\documentclass[a4paper,fleqn,usenatbib]{mnras}

\usepackage{newtxtext,newtxmath}

\usepackage[T1]{fontenc}
\usepackage{ae,aecompl}

\usepackage{graphicx}	
\usepackage{amsmath}	
\usepackage{amssymb}	

\newcommand{\kms}{\ensuremath{\mathrm{km}\,\mathrm{s}^{-1}}}

\newcommand{\MLsun}{\ensuremath{\mathrm{M}_{\odot}/\mathrm{L}_{\odot}}}

\newcommand{\Msun}{\ensuremath{\mathrm{M}_{\odot}}}

\newcommand{\ML}{\ensuremath{\Upsilon_*}}
\newcommand{\Mst}{\ensuremath{M_*}}

\title[MOND and NGC1052-DF2]{MOND and the dynamics of NGC1052-DF2}

\author[B. Famaey, S. McGaugh, M. Milgrom]{
B. Famaey,$^{1}$\thanks{E-mail: benoit.famaey@astro.unistra.fr stacy.mcgaugh@case.edu moti.milgrom@weizmann.ac.il}
S. McGaugh,$^{2}$
and M. Milgrom$^{3}$
\\
$^{1}$Universit\'e de Strasbourg, CNRS  UMR  7550,  Observatoire  astronomique  de  Strasbourg, 67000  Strasbourg,  France\\
$^{2}$Department of Astronomy, Case Western Reserve University, Cleveland, OH 44106, USA\\
$^{3}$Department of Particle Physics and Astrophysics,
 Weizmann Institute of Science, Rehovot 76100, Israel
}

\date{Accepted XXX. Received YYY; in original form ZZZ}

\pubyear{2018}

\begin{document}
\label{firstpage}
\pagerange{\pageref{firstpage}--\pageref{lastpage}}
\maketitle

\begin{abstract}
The dwarf galaxy NGC1052-DF2 has recently been identified as potentially lacking dark matter. If correct, this could be a challenge for MOND, which predicts that low surface brightness galaxies should evince large mass discrepancies. However, the correct prediction of MOND depends on both the internal field of the dwarf and the external field caused by its proximity to the giant elliptical NGC1052. Taking both into consideration under plausible assumptions, we find $\sigma_{\rm MOND} = 13.4^{+4.8}_{-3.7}\;\kms$. This is only marginally higher than the claimed 90\% upper limit on the velocity dispersion ($\sigma < 10.5\;\kms$), and compares well with the observed root mean square velocity dispersion ($\sigma = 14.3\;\kms$). We also discuss a few caveats on both the observational and theoretical side. On the theory side, the internal virialization time in this dwarf may be longer that the time scale of variation of the external field. 
On the observational side, the paucity of data and their large uncertainties call for further analysis of the velocity dispersion of NGC1052-DF2, to check whether it poses a challenge to MOND or is a success thereof.
\end{abstract}




\section{Introduction}

In a recent paper, \citet{vanDokkum} report the line-of-sight velocities of ten globular clusters around the dwarf galaxy NGC1052-DF2. They infer from these a surprisingly low velocity dispersion for such a galaxy: $\sigma < 10.5\;\kms$ at 90\% confidence. They then deduce that the total mass within the outermost radius of 7.6 kpc would be $< 3.4 \times 10^8 \, \Msun$. The stellar mass which \citet{vanDokkum} assume for this galaxy, $\Mst \approx 2 \times 10^8\;\Msun$, is based on a stellar mass-to-light ratio of $\ML \approx 2\;\MLsun$. This stellar mass is close to the dynamical mass, implying no need for dark matter. If all these deductions are correct, this would in principle contradict MOND.

MOND \citep{Milgrom1983,FamaeyMcG,2014SchpJ} is a paradigm predicting the dynamics of galaxies directly from their baryonic mass distribution. The postulate is that, for gravitational accelerations  below $a_0 \approx 10^{-10} {\rm m} \, {\rm s}^{-2}$ the actual gravitational attraction approaches $(g_N a_0)^{1/2}$ where $g_N$ is the usual Newtonian gravitational attraction.

The successes of MOND are best known for galaxies that are isolated, axisymmetric, and rotationally supported \citep[e.g.,][]{SM02,Gentile}. In such systems, there is a clear and direct connection between the baryonic mass distribution and the rotation curve \citep{MLS,OneLaw}. This empirical relation is indistinguishable from MOND \citep{Lifits}.

MOND also has a good track record of predictive success for pressure supported systems like NGC1052-DF2 \citep{MM13a,MM13b,PM14,Crater2pred}. The analysis in such cases is complicated by the same uncertainties as in Newtonian analyses, such as that in the stellar mass-to-light ratio and the unknown anisotropy in the velocity tracers. Unique to MOND is the external field effect \citep[EFE,][]{Milgrom1983, BM84, FamaeyMcG, Haghi16, Hees}. Because of the nonlinearity of MOND, the internal dynamics of a system can be affected by the external gravitational field in which it is immersed. When the external field dominates over the internal one, the amplitude of the MOND effect, and the corresponding amount of dark matter inferred, is reduced. Some interesting effects unique to MOND are related to the EFE, such as the prediction of asymmetric tidal streams of globular clusters \citep{Thomas}.

An essential consequence of the EFE in MOND is that the predicted velocity dispersion of a dwarf galaxy depends on its environment. An object in isolation is expected to have a higher velocity dispersion than the same object in orbit around a massive host. This difference is perceptible in pairs of photometrically indistinguishable dwarf satellites of Andromeda \citep{MM13b}. Indeed, the EFE was essential to the correct \textit{a priori} prediction \citep{MM13a} of the velocity dispersions of the dwarfs And XIX, And XXI, and And XXV. These cases are notable for their large scale lengths and low velocity dispersions \citep{Collins2013} --- properties that were surprising in the context of dark matter but are natural in MOND. A further example is provided by the recently discovered Milky Way satellite Crater 2 \citep{Crater2discovery}. \citet{Crater2pred} predicted that this object would have a velocity dispersion of $2.1^{+0.9}_{-0.6}\;\kms$, much lower than the nominal expectation in the context of dark matter. \citet{Crater2vdisp} subsequently observed $2.7\pm0.3\;\kms$.

 Here we take the external field of the host galaxy NGC1052 into account to predict the expected velocity dispersion of NGC1052-DF2 in MOND. In \S   \ref{sec:prediction} we adopt a nominal mass-to-light ratio of $\ML = 2\;\MLsun$ to obtain $\sigma_{\rm MOND} \approx 13.4\;\kms$. For a factor of two uncertainty in the $V$-band stellar mass-to-light ratio, the allowed range is $9.7 \leq \sigma_{\rm MOND} \leq 18.2\;\kms$. This range of variation follows simply from plausible uncertainty in the conversion of luminosity to stellar mass; other potentially relevant uncertainties are discussed further in \S \ref{sec:caveats}. Brief conclusions are given in \S \ref{sec:conc}.

\section{Predicting the Velocity Dispersion of NGC1052-DF2 in MOND}
\label{sec:prediction}

\subsection{Isolated prediction}

\citet{MM13a} outlined the procedure by which the velocity dispersions of satellite galaxies like NGC1052-DF2 can be predicted. For a spherical, isotropic, isolated system, 
\begin{equation}
\sigma_{\rm iso} = \left(\frac{4}{81} G M a_0 \right)^{1/4} \simeq \left(\frac{\ML}{2}\right)^{1/4} 20\;\kms\;\mathrm{for~NGC1052\text{-}DF2.}
\end{equation}
This is the same result obtained by \citet{vanDokkum} for the same assumptions. However, this is not the correct MOND prediction, as this dwarf is not isolated: the external field due to the giant host NGC1052 is not negligible. 

\subsection{Instantaneous external field effect prediction}

The isolated MOND boost to Newtonian gravity is predicted to be observed only in systems where the \textit{absolute} value of the gravity both internal, $g$, and external, $g_e$ (from a host galaxy, or large scale structure), is less than $a_0$ . If $g_e < g < a_0$ then we have isolated MOND effects. 
If instead $g < g_e < a_0$, then the system is quasi-Newtonian. The usual Newtonian formula applies, but with a renormalized gravitational constant. When $g_e \ll a_0$, the renormalizing factor is simply $a_0/g_e$. 

The velocity dispersion estimator is simple \citep{MM13a} when $g \gg g_e$ (isolated) or $g \ll g_e$ (EFE) in that it depends only on the dominant acceleration ($g$ or $g_e$). When $g \approx g_e$, both must be taken into consideration. This turns out to be the case for NGC1052DF2.

To estimate these quantities, we make the same assumptions as \citet{vanDokkum}. For a $V$-band stellar mass-to-light ratio of $\ML \simeq 2$, the baryonic mass is $2 \times 10^8\;\Msun$. At the deprojected 3D half-light radius $r_{1/2} = (4/3)R_e = 2.9 \, {\rm kpc}$, the internal Newtonian acceleration is $g_N = 1.34 \times 10^{-2} a_0$. This corresponds to $g_i \sim 0.12 a_0$ for an isolated object in MOND. 

For the external field, we assume $g_e = V^2/D$ assuming a flat rotation curve for the host NGC1052 of $V = 210\;\kms$, in accordance with the stellar mass of the host estimated by \citet{SLUGGS} and consistent with the kinematic observations of \citet{vG86}. At the projected distance of NGC1052-DF2 from its host, this yields an external field $g_e = 0.15 a_0$. This estimate of the external acceleration is rather uncertain for observational, not theoretical, reasons. For specificity, we adopt $g_e = 0.15 a_0$ here, and discuss the uncertainties further in \S \ref{sec:caveats}.

The internal isolated MOND acceleration $g_i \sim 0.12 a_0$ and the external one from the host $g_e \sim 0.15 a_0$ are thus of the same order of magnitude. The exact calculation should then be made with a numerical Poisson solver. Nevertheless, a good ansatz in such a case is to consider the net MOND effect in one dimension \citep[eq. 59 of][]{FamaeyMcG}:
\begin{equation}
(g + g_e) \, \mu\left(\frac{g+g_e}{a_0}\right) = g_N + g_e \mu \left( \frac{g_e}{a_0} \right) ,
\label{efe}
\end{equation}
Here, $\mu$ is the MOND interpolating function, $g$ the norm of the internal gravitational field one is looking for, $g_e$ the external field from the host, and $g_N$ the Newtonian internal gravitational field. Adopting here the `simple' interpolating function\footnote{The result is not particularly sensitive to the choice of interpolation function since the object in question is deep in the regime of low acceleration where all interpolation functions converge.} of MOND which is known to provide good fits to galaxy rotation curves \citep{FamaeyB, Gentile}, this equation can easily be solved for $g$. \citet{Wolf} provides us with a mass estimator at the deprojected 3D half-light radius $r_{1/2}$, where we can consider that the system feels the effect of the renormalized gravitational constant in MOND, $G_{\rm eff} = G \times [g(r_{1/2})/g_N(r_{1/2})]$. 

Solving Eq.~(2) yields $G_{\rm eff} = 3.64 G$ at the half-light radius. Using this and solving the mass estimator of \citet{Wolf} for the velocity dispersion leads to
\begin{equation}
\sigma_{\rm MOND} \approx 13.4 \, {\rm km} \, {\rm s}^{-1}. 
\end{equation}
This is lower than the isolated case ($20\;\kms$) considered by \citet{vanDokkum}, as is always the case when the EFE is important. This analytic estimate of the predicted velocity dispersion in MOND is in good agreement with an independent estimate by \citet{Kroupa}, whose work is based on fits to numerical simulations by \citet{Haghi} of globular clusters embedded in an external field.

To estimate an uncertainty on this prediction, we consider a factor of two variation in the $V$-band mass-to-light ratio. That is, we obtain $\sigma_{\rm MOND}=9.7\;\kms$ for $\ML = 1\;\MLsun$, and $\sigma_{\rm MOND}=18.2\;\kms$ for $\ML = 4\;\MLsun$. Whether this range is appropriate for this particular dwarf is pure supposition that we must make in any theory. Note also that if we apply the same range to the mass of the host galaxy, this largely compensates for possible variations in distance of NGC1052-DF2 with respect to the host.

The MOND velocity dispersion discussed by \citet{vanDokkum}, $20\;\kms$, only applies in isolation. It simply is not the correct prediction for MOND in the case of NGC1052-DF2 given its proximity to the giant elliptical NGC1052. The velocity dispersion we predict for MOND with proper consideration of the EFE disagrees far less with the data of \citet{vanDokkum} than the isolated case. Rather than being removed from the 90\% c.l.\ upper limit of $10.5\;\kms$ by a factor of two, the MOND prediction differs by only a few \kms\ and overlaps with it at the $1 \sigma$ level. The correct MOND prediction is very close to the raw r.m.s. velocity dispersion ($14.3\;\kms$) measured by \citet{vanDokkum}. Clearly we cannot exclude the theory when one realization of the data is entirely consistent with it.

\section{Caveats}
\label{sec:caveats}

The prediction of velocity dispersions in MOND is most clean and simple in the isolated, low acceleration case, depending only on the stellar mass given the usual assumptions of spherical symmetry, isotropy, and dynamical equilibrium. Examples of isolated dwarfs are provided by And XXVIII \citep{MM13a,MM13b} and Cetus \citep{PM14}. The problem becomes more involved when the EFE is relevant. In addition to the stellar mass of the dwarf, the extent of the stellar distribution matters to the determination of the internal acceleration. Diffuse objects\footnote{The quantity of merit in MOND is the surface density, not the mass. Lower surface density means lower acceleration. That NGC1052-DF2 is massive for a dwarf does not shield it from the EFE as it is a diffuse object.} like NGC1052-DF2 or Crater 2 are more subject to the EFE than compact objects like globular clusters. In addition to the internal field, we also need to measure the external field. This depends on the mass of the host and the 3D distance separating the host from its dwarf satellite. In dense environments where there may be multiple massive systems, matters obviously become more complicated still.

Estimating the external field can be challenging. We discuss here a few of the issues that arise, both observational (how well do the data constrain $g_e$) and theoretical (is it fair to assume the instantaneous value of the EFE at this moment in the dwarf's orbit). The estimates made above adopt reasonable assumptions consistent with those of \citet{vanDokkum}, but these need not be correct.

\subsection{Observational caveats}

We have already noted that the r.m.s velocity dispersion ($14.3\;\kms$) is higher than the 90\% c.l.\ upper limit ($10.5\;\kms$). This odd situation stems entirely from the omission of a single velocity tracer by \citet[the object labeled 98 in their Fig.~2]{vanDokkum}. While this is an outlier statistically, its exclusion is a choice that leads to a perverse physical situation. Without this one globular cluster, the mass of the object is too low for it to remain gravitationally bound. This raises the question as to why it is present at all, which is a essentially the original puzzle pointed out by \citet{Zwicky}. Stranger still, once excluded, \citet{vanDokkum} argue that the intrinsic velocity dispersion is as low as $3.2\;\kms$, at which points there is not enough dynamical mass to explain the observed stars. It is beyond the scope of this work to reanalyze these data; hopefully others will comment independently on the probability distribution of the velocity dispersion when only ten velocity tracers are available. One might also be concerned whether this tracer population is representative of the mass-weighted velocity dispersion \citep[see discussion in][and references therein]{QUMOND}, or if nearly face-on rotation of the globular cluster system could play a role.

A key uncertainty in our analysis is the distance, which is required to determine the absolute acceleration scale. Distances are notoriously difficult to constrain for individual galaxies. We largely agree with the assessment of the distance laid out by \citet{vanDokkum}, but for completeness point out where this matters most.

The absolute distance to NGC1052-DF2 matters insofar as this determines its association with the putative giant host NGC1052. If the two are not associated, then the object is likely isolated. Even that is not guaranteed given the proximity of NGC1042 along the line of sight. Setting that aside, a closer distance would be more consistent with the isolated MOND case as the stellar mass shrinks as $D^2$ and the predicted velocity dispersion along with it. Conversely, a significantly larger distance (beyond NGC1052) would be problematic for MOND as the object would then be more massive and should have a higher velocity dispersion.

The absolute 3D distance relative to the host NGC1052 also matters, as it affects the strength of the external field. While the projected separation is well determined for the assumed distance, the additional distance along the line of sight is not. As this increases, the EFE weakens, tweaking the calculation above. However, this is largely degenerate with uncertainties in the mass of the host, so it is difficult to provide a sensible error estimate that does not look like hedging. These are observational uncertainties, not theoretical hedges.

\subsection{Theoretical caveats}

MOND is a nonlinear theory. The approximation we have adopted to predict the velocity dispersion implicitly assumes that the external field is effectively constant. In reality, the amplitude and vector direction of the external field varies as the dwarf orbits its host. The issue then becomes whether the internal dynamics of the dwarf have time to come into equilibrium with the continually changing external field \citep{bradamilgrom}. If so, the prediction we obtain is valid. If not, it is no more valid than if the condition of dynamical equilibrium is not obtained in Newtonian dynamics, as would be the case if the dwarf were being tidally stripped of its dark matter.

In MOND, the internal dynamics of an object depends on the external field in which it is immersed as soon as the external field dominates over the internal one. If the external field does not vary quickly in time with respect to the time it takes for the object to settle back to virial equilibrium, the ``instantaneous" external field approximation that we have applied is appropriate. This works well for quasi-circular orbits, as the amplitude of the external field is steady even though the orientation vector varies. In general, the external field is expected to change over time scales $D/V$, where $D$ is the distance to NGC1052 and $V$ its circular velocity. This timescale is of the order of $3 \times 10^8$ years. The relevant internal time is the time it takes for the system (in the present case the system of globular clusters being utilized as velocity tracers) to come back to virial equilibrium. If this would be just half a revolution around NGC1052-DF2  at a typical radius of 5 kpc for the globular clusters, and for a typical velocity of $10 \, {\rm km} \, {\rm s}^{-1}$, this time would be $1.6 \times 10^9$ years. So the dynamical state of the system of globular clusters has not obviously had time to equilibrate. Consequently, it might retain a memory of a previous value of the external field, and the system would then basically be out-of-equilibrium. Much depends on the details of its orbit: whether it is on a radial or circular orbit, and when its last pericenter passage was. One should bear in mind that the same issues pertain to tidal disruption in a Newtonian analysis, albeit with a different amplitude: one obvious interpretation for a dwarf to be devoid of dark matter is that it has been tidally stripped thereof, in which case it is not in equilibrium and any velocity tracers must be interpreted with a grain of salt.

The stability of dwarfs orbiting massive hosts has been investigated numerically by \citet{bradamilgrom}. Applying their criteria as in \citet{McGWolf}, it indeed appears that NGC1052-DF2 may be on the fringe where dynamical equilibrium ceases to apply for the ultrafaint dwarfs of the Milky Way. More generally, \citet{bradamilgrom} describe how an orbit dwarf may experience oscillations in velocity dispersion and scale length as it orbits from pericenter to apocenter. It is easy to imagine that NGC1052-DF2 is in a phase where its velocity dispersion is minimized while the scale length is maximized. While there is no way to ascertain if this is the case, this is an anticipated effect in MOND, and illustrates some of the subtle nonlinear effects that can arise.

We may also estimate the tidal radius in MOND. At the projected distance from the host, this is $\sim 8$ kpc. Intriguingly, this is just beyond the outermost globular cluster. Consequently, the limit on the distance from the host obtained from the Jacobi radius by \citet{vanDokkum} does not apply in MOND.

The caveats discussed in this subsection are theoretical: nonlinear theories can be sensitive, and great care must be taken to extract the correct prediction. That said, we see no reason to expect the bulk velocity dispersion to vary outside the bounds stipulated by the plausible range in the stellar mass-to-light ratio, at least for our calculations referenced to the half-light radius. It would be interesting to compute the detailed radial variation of the velocity dispersion \citep{Alexander}, but such a computation is well beyond the scope of this letter, and indeed, beyond the scope of foreseeable observations to test for this remote dwarf galaxy.

\section{Conclusions}
\label{sec:conc}

We have predicted the velocity dispersion of the dwarf galaxy NGC1052-DF2 in MOND. We find $\sigma_{\rm MOND} \approx 13.4^{+4.8}_{-3.7}\;\kms$, not the $\sim 20\;\kms$ discussed by \citet{vanDokkum}. The difference arises because of the external field effect. The acceleration of stars within the dwarf is comparable to the acceleration of the dwarf in its orbit around its host galaxy, NGC1052. Consequently, neither field may be ignored, and the net MOND effect is reduced from the isolated case. This, in turn, reduces the predicted velocity dispersion, which consequently is not in conflict with the measurements of \citet{vanDokkum}.





\bsp	
\label{lastpage}
\end{document}